\documentclass[showpacs,preprintnumbers,prc]{revtex4}
\usepackage{color}
\usepackage{graphicx}
\newcommand{\be}{\begin{equation}}
\newcommand{\ee}{\end{equation}}
\newcommand{\bea}{\begin{eqnarray}}
\newcommand{\eea}{\end{eqnarray}}
\newcommand{\nn}{\nonumber}

\newcommand{\sls}{\!\!\!/}
\newcommand{\noi}{\noindent}

\begin{document}
\title{\bf{Drag and diffusion coefficients of heavy quarks in hard thermal loop approximations}}
\author{Trambak Bhattacharyya\footnote{trambakb@vecc.gov.in}, Surasree 
Mazumder and Jan-e Alam}
\medskip
\affiliation{Theoretical Physics Division, 
Variable Energy Cyclotron Centre, 1/AF, Bidhan Nagar, Kolkata-700064}
\date{\today}
\begin{abstract}
The drag and  diffusion coefficients of heavy quarks propagating through quark gluon plasma
(QGP) have been evaluated using Hard Thermal Loop (HTL) approximations. The HTL corrections
to the relevant propagators and vertices have been considered. 
It is observed that the magnitudes of both the transport coefficients 
are changed significantly from values obtained by earlier approaches where
either (i) the $t$ channel divergence in $T=0$ pQCD matrix element is shielded simply by Debye mass. 
or (ii) only HTL resummed propagator is used ignoring the HTL corrections at the interaction vertices.
The implications of these changes in the transport coefficients on the heavy ion phenomenology have been discussed. 
\end{abstract}
\maketitle

\section{Introduction}
The study of the transport coefficients of strongly coupled system is a field of
high contemporary interest both theoretically and experimentally. In one hand, the calculation of
the lower bound on the shear viscosity ($\eta$) to entropy density ($s$)  ratio ($\eta/s$) within the 
frame work of AdS/CFT model~\cite{kss}  has ignited enormous interests among the theorists. On the 
other hand, the experimental study of the $\eta/s$ for cold atomic systems and QGP 
and their similarities have generated huge interest across various branches of 
physics (see~\cite{aadams} for a review).
In general, the interaction of probes with a medium  brings out useful information about the
nature of the medium.  Since the magnitude of the transport coefficients are sensitive to 
the coupling strength, hence these quantities can be adapted as useful quantities to characterize 
a  medium. In the context of probing QGP, expected to be produced in ultra-relativistic
heavy ion collisions at Relativistic Heavy Ion Collider (RHIC) and Large Hadron Collider (LHC) 
energies, we choose the heavy quarks (HQs), charm and beauty, as probes. That is, we would like to
extract the drag and diffusion coefficients of the QGP by studying the propagation of HQs through
QGP.  Selection of HQs as probes has several advantages, such as (i) they are produced very early in the
collisions and remain extant throughout the evolution of the QGP. As a result, the HQs witness
the entire evolution of the system. 
It is expected that the HQ thermalization time
is larger than the light quarks by a factor $m/T$ where $m$ is the mass of the HQ and $T$ is 
the temperature.  Therefore,  the HQs may remain out of equilibrium in QGP.
(ii)The chances of HQs getting thermalized in the system 
is weaker and hence do not dictate the bulk properties of QGP.  
Moreover, the  observed transverse momentum suppression ($R_{AA}$) of 
leptons originating from the decays of heavy flavours produced in nuclear 
collisions as compared to those produced in proton+proton (pp) collisions at
the same colliding energy~\cite{stare,phenixe,alice} offer us an opportunity to estimate
the drag and diffusion coefficients of QGP.  

In the present circumstances a description of the motion of the non-equilibrated HQs 
in the background of equilibrated system of QGP is required. 
An appropriate foundation is provided by 
the Fokker-Planck (FP) equation~\cite{landau,balescu}, which reads as follows:
\bea
\frac{\partial{f}}{\partial{t}}=\frac{\partial}{\partial{p_i}}
\left[A_i(p)f+\frac{\partial}{\partial{p_i}}[B_{ij}f]\right]
\label{eqn1}
\eea
where $f$ stands for the momentum-space distribution of the particle (HQs here) undergoing Brownian
motion in the thermal bath of QGP. The question of HQ thermalization can be addressed by comparing the
solution of FP equation with the HQ's thermal distribution at any given time.  
$A_i$ and $B_{ij}$ of Eq.~\ref{eqn1} are related to the drag and diffusion 
coefficients respectively. Hence, the interactions of the HQs  with the QGP  are incorporated in  
$A_i$ and $B_{ij}$. That means, $A_i$ and $B_{ij}$ can supply information
about the nature of the QGP~\cite{sc,rapp,turbide,bjoraker,sve,japrl,
npa1997,munshi,rma,mbad,santosh,moore}.
The issue of HQ thermalization in QGP can also be addressed experimentally 
by measuring the elliptic flow ($v_2$) of leptons from the decays
of HQs.  Therefore, the evaluation of the 
drag and diffusion coefficients of QGP become extremely important.  
We will see below that drag (diffusion) coefficients are, essentially, 
momentum (square of the momentum) transfer 
weighted over the squared interaction matrix element ($|M|^2$). 
This indicates that  an accurate evaluation of $|M|^2$ is of vital importance.
In the present work we attempt to estimate both these coefficients
by using the techniques of thermal field theory in hard thermal loop (HTL)
approximations using resummed gluon propagator and one loop corrections to relevant vertices.
Some of the earlier attempts \cite{heisel,qunwang} lack the vertex corrections which is necessary for maintaining 
gauge-invariance. 

The two main elastic processes which contribute to the transport coefficients
are: $Q+q\rightarrow Q+q$ and  $Q+g\rightarrow Q+g$. Here $Q$ ($q$) stands for heavy (light) quarks
and $g$ denotes gluon. The 
$|M|^2$ for these processes contain $t$-channel divergence which are normally 
regulated by introducing thermal mass ($m_D$) for the exchanged gluons~\cite{mbad,sve} i.e. by 
replacing $t$ by $t-m_D^2$ in the denominator of the matrix elements. 
In the present work,  instead of shielding the divergences simply by (static) Debye mass 
we will  use the  HTL approximated gluon propagator in the $t$-channel diagrams with vertex correction
in a self-consistent way.

The paper is organized as follows. In the next section the general 
expressions for the drag and diffusion
coefficients are outlined.  In section III we briefly discuss the effective
gluon propagators in HTL approximations. The significance of the effective three gluon ($ggg$)
and quark-gluon ($qqg$) vertices correction is discussed in the context of gauge invariance.
Section IV is devoted for presenting results on the
drag and diffusion coefficients. The summary and conclusions of
the present work is presented in section V. The appendix contains the detailed derivation of the
matrix elements required for the evaluation of drag and diffusion coefficients.

\section{The drag and diffusion coefficients}
In terms of the transition rates the collision integral of the Boltzmann transport equation can be written as~\cite{landau}:
\be
\left[\frac{\partial f}{\partial t}\right]_{collisions}= \int d^{3}\textbf{k}[w(\textbf{p}+\textbf{k},\textbf{k})
f(\textbf{p}+\textbf{k})-w(\textbf{p},\textbf{k})f(\textbf{p})].
\label{eqn2}
\ee
where $w(\textbf{p},\textbf{k})$ is the collision rate, say for the processes,  ${\bf Q(p)+g(q)\rightarrow Q(p-k)+g(q+k)}$,
where the quantities within the bracket denotes the corresponding momenta of the particle.
Using Landau approximation {\it i.e.} by expanding  
$w(\textbf{p}+\textbf{k},\textbf{k})$ in powers of  $\textbf{k}$
and keeping upto quadratic term,  
the Boltzmann transport equation can be written as~\cite{mbad,sve}
\be
\frac{\partial f}{\partial t}= \frac{\partial}{\partial p_{i}}\left[A_{i}(\textbf{p})f+\frac{\partial}{\partial p_{j}}[B_{ij}
(\textbf{p})f]\right]~~,
\label{eqn3}
\ee
where the kernels are defined as
\be
A_{i}= \int d^{3}\textbf{k}w(\textbf{p},\textbf{k})k_{i}~~,
\label{eqn4}
\ee
and
\be
B_{ij}= \frac{1}{2} \int d^{3}\textbf{k}w(\textbf{p},\textbf{k})k_{i}k_{j}.
\label{eqn5}
\ee
For $\mid\bf{p}\mid\rightarrow 0$,  $A_i\rightarrow \gamma p_i$
and $B_{ij}\rightarrow D\delta_{ij}$ where $\gamma$ and $D$ stand for
drag and diffusion coefficients respectively. The drag and diffusion coefficients have
recently been evaluated within the ambit of AdS/CFT~\cite{ssgubser} and pQCD~\cite{huot}
and their importance for jet quenching have been discussed. 
Eq.~\ref{eqn3} is a nonlinear integro-differential equation known as the Landau kinetic equation. 
The appearance of parton distribution in the expression for $\omega$ makes Eq.~\ref{eqn3} a
non-linear one. 
For the problem under consideration one of the colliding partners (light quarks or gluons) 
is in equilibrium. In such a situation
the distribution function which appears in $w$ can be replaced by thermal distribution. As a 
consequence Eq.\ref{eqn3} becomes a linear partial differential equation, known as Fokker-Planck (FP) equation.
The $T$ dependence of the transport coefficients enter through the thermal distribution appearing in $\omega$.

As mentioned above the drag and diffusion coefficients are related to the quantities
$A_i$ and $B_{ij}$. Both these coefficients can be calculated from the following expression~\cite{sve} with
appropriate choice of the function $F(p^\prime)$,

\bea
\langle\langle F(p)\rangle\rangle&=&\frac{1}{512\pi^4}\frac{1}{E_p}
\int_0^{\infty}q dq d(cos\chi)\frac{s-m^2}{s}f(q)\int_{-1}^1 \nn\\
&&d(cos\theta_{c.m.}) \frac{1}{g_Q}\overline{|M|}^2
\int_0^{2\pi}d(\phi_{c.m.}) F(p^{\prime})
\label{eqn6}
\eea
where $g_Q$ is the HQ degeneracy,  $F(p^{\prime}=p-k)$ is a function of $p,~q$ and CM frame scattering angles and $cos\chi$ can be 
obtained from,
\be
s=p^2+q^2+2(E_pE_q-|\vec{p}||\vec{q}|cos\chi)
\label{eqn7}
\ee 
$\theta_{c.m.}$ and $\phi_{c.m.}$ are polar and azimuthal angles of q respectively.
Drag ($\gamma$) can be obtained by the following replacement in  Eq.~\ref{eqn6}:
\bea
F(p^{\prime})=1-\frac{p.p^{\prime}}{p^2} 
\label{eqn8}
\eea
For determining diffusion ($D$) we substitute,
\bea
F(p^\prime)=\frac{1}{4}\left[p^{\prime2}-\frac{(p.p^{\prime})^2}{p^2}\right]
\label{eqn9}
\eea
in Eq.~\ref{eqn6}.

\section{Resummed gluon propagator, effective three gluon and quark-quark-gluon vertices in HTL approximation}
As discussed before the calculation of drag and diffusion coefficients
involve the evaluation of amplitudes for processes like $Q+q\rightarrow Q+q$ and $Q+g\rightarrow Q+g$ ~\cite{combridge}.
The amplitudes from bare perturbation theory contains $t$-channel divergence due to low four-momentum, 
$P =(\omega,\vec{p})$ gluon exchange. This divergence can be regulated by introducing thermal 
mass of gluon. Here,  we study the HTL approximations~\cite{BP} and resummation of gluon propagator 
which will enable us to regulate the $t$-channel divergence in a self-consistent
way and hence will lead to comparatively more reliable values of the transport coefficients. 

\begin{figure}[h]      
\begin{center}                                                               
\includegraphics[scale=0.9]{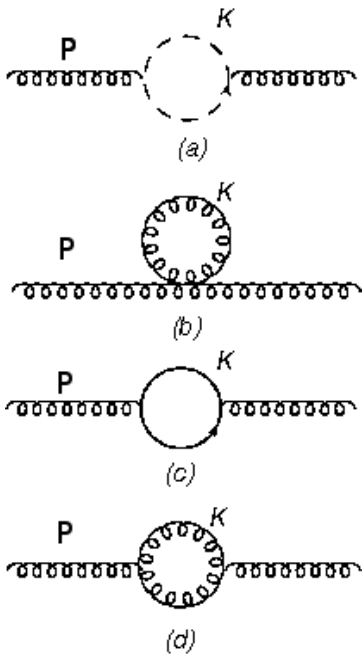}                                 
\end{center}                                                                 
\caption{Feynman diagrams contributing to gluon self-energy upto one loop. (a)ghost-gluon loop. 
(b)four-gluon vertex. (c) quark-antiquark pair creation. (d)three-gluon vertex.}
\label{fig1}                                                               
\end{figure}

Our aim is to find out HTL approximated self-energy of gluon which goes as an input to the resummed 
gluon propagator to be used as effective thermal propagator regularizing the $t$ channel divergence.
The gluon self-energy in HTL approximation is discussed in detail in Ref.~\cite{bellac,kap}. 
In this section we  give only an outline of the scheme. 
There are four diagrams which contribute to gluon  self-energy~(Fig.~\ref{fig1}).
The loop integrations can be written down easily if we keep in mind that the loop-momentum, $K=(k_0,\vec{k})$
is `hard' compared to external gluon momentum, {\it i.e.} $P>>Q$ which
enables us to use simplified $ggg$ vertex \cite{bellac}. Our goal will be to find out $T^2$
contributions of self-energy because the momentum integration is cut-off at
the momentum scale$\sim T$ due to the presence of thermal distribution function. 
That is we can take-up the momentum integration $\int_0^{\infty} k dk$ which blows
 up as $k\rightarrow \infty$, but, insertion of thermal distribution function makes it finite 
even at  $k\rightarrow \infty$,

\be
\int_0^{\infty} k f(k) dk=\frac{\pi^2 T^2}{12}
\label{eqn10}
\ee
The leading contribution in Eq. \ref{eqn10} is given by $k\sim T$.
If we are interested in high-temperature limit we can assume $|\vec{k}|>>|\vec{p}|$ and approximate 
related quantities accordingly~\cite{bellac}. 

The effective gluon propagators evaluated in one loop order in HTL approximation
enter in the transport coefficients evaluated for the processes displayed 
in Figs.~\ref{fig8} and ~\ref{fig11}. For low momentum transfer (i.e.$\sim gT$ where $g$ is 
the colour charge, $g=\sqrt{4\pi \alpha_s(T)}$ and $\alpha_s(T)$ is the strong coupling), 
one has to use the resummed propagator~\cite{heisel}. The resummed gluon propagator, which is given by: 
\bea
\Delta^{\mu\nu}=\frac{\mathcal{P}_T^{\mu\nu}}{-P^2+\Pi_T}+\frac{\mathcal{P}_L^{\mu\nu}}{-P^2+\Pi_L}
+(\alpha-1)\frac{P^{\mu}P^{\nu}}{P^2}
\label{eqn11}
\eea
will need HTL approximated $\Pi_L$ and $\Pi_T$. 
The transverse and longitudinal self-energies,  $\Pi_T$ and $\Pi_L$ are given by
\bea
\Pi_L(P)=(1-x^2)\pi_L(x),~~~~\Pi_T(P)=\pi_T(x)
\label{eqn12}
\eea
where $x=\omega/q$ and scaled self-energies $\pi_T$ and $\pi_L$ are given by~\cite{bellac}, 

\bea
\pi_T(x)=
m_D^2\left[\frac{x^2}{2}+\frac{x}{4}(1-x^2)ln \left(\frac{1+x}{1-x}\right) 
-i\frac{\pi}{4}x(1-x^2)\right]\nn\\
\eea
\bea
\pi_L(x)=m_D^2\left[1-\frac{x}{2}ln (\frac{1+x}{1-x})+i \frac{\pi}{2}x\right]\nn\\
\label{eqn13}
\eea
where $m_D$ in Eq. \ref{eqn13} is the thermal mass of gluon; and
is given by $m_D^2=g^2T^2(C_A+N_f/2)/6$,
we use 2-loop perturbative temperature-dependent coupling for our calculation \cite{kacz}
{\it i.e.} in the present calculations the strong coupling runs with temperature.   
In addition to the HTL corrections to the propagator 
we  introduce the HTL corrections to the $ggg$ and $qqg$ vertices in t-channel diagram of $Qg\rightarrow Qg$ process
and $Qq\rightarrow Qq$ processes
to maintain gauge-invariance \cite{brateenprl}. The involvement of heavy quarks in the dynamical processes
helps us to 
approximate heavy quark (HQ) propagators and $HQ-HQ-g$ vertex by their $T=0$ counterparts. But
the matrix elements under consideration have their origin from cutting through the heavy quark 
self-energy diagram using effective gluon propagator. Hence inclusion of effective $qqg$
and $ggg$ vertex becomes inevitable. Some recent works \cite{rapphq} refrain from using effective 
vertices because the gluons and light quarks are `hard' as they are in thermal bath;
and following the argument of \cite{bellac}, the uncorrected $ggg$ vertex ($\mathcal{O}(T)$), then,
dominates over corresponding HTL vertex correction ($\mathcal{O}(g^3 T)$). So, the vertex-corrections can
approximately be neglected though one should maintain gauge-invariance by 
as gauge-symmetry is a sacred symmetry of the strong interaction.

From the previous discussion we have seen that the leading behaviour in temperature of
gauge particle self-energies is proportional to $T^2$. This result can be generalized to
$N$-point functions, computed at the one-loop approximation. The HTL corrections
to $ggg$ and $qqg$ vertices can be obtained from \cite{bellac,anderson}.
(see Appendix for detailed discussion)

\section{Results}
The drag and diffusion coefficients of HQ, propagating through QGP and suffering elastic collisions, 
evaluated by using the HTL approximated effective gluon propagators and effective $ggg$ and $qqg$ vertices
are  denoted by $\gamma_{HTL}$ and $D_{HTL}$ respectively. 
To present the results of our calculations we use the following notations.
The drag and diffusion coefficients evaluated with bare vertices and propagators 
will be denoted by $\gamma$ and $D$ respectively (the Debye mass is introduced in the
t-channel propagator to shield the infra-red divergences). 
The $\gamma_{HTL}$ and $D_{HTL}$ 
are calculated with the HTL approximated propagator and vertices. 
All these quantities, {\it i.e.} 
 $\gamma$, $D$,  $\gamma_{HTL}$ and $D_{HTL}$  are evaluated 
 with the same kinematic approximations~\cite{bellac} to make the comparison of
the bare and HTL approximated quantities meaningful.
 
\begin{figure}[h]      
\begin{center}                                                               
\includegraphics[scale=0.5]{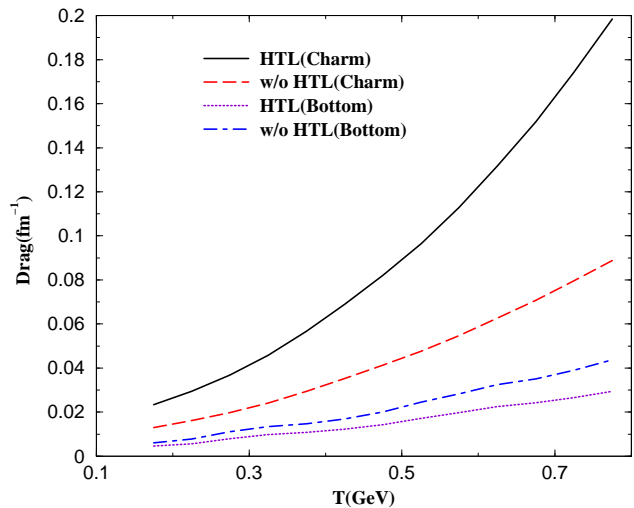}                                 
\end{center}                                                                 
\caption{ (Color online) Variation of drag of heavy quarks 
of momentum 1 GeV with temperature.}
\label{fig2}                                                               
\end{figure}

The variations of drags with temperature for HQs at momentum, $p=1$ GeV 
are displayed in Fig.~\ref{fig2}. The results
clearly indicate an enhancement and  a more rapid variation of $\gamma_{HTL}$ compared to $\gamma$.
The increase is more prominent for charm than beauty.   
We have explicitly checked that in the static limit ($\omega/q=x\rightarrow 0$) 
the $\gamma_{HTL}$ approaches $\gamma$. 
Results displayed in Fig. \ref{fig2} indicate that at $T=400$ MeV the  $\gamma_{HTL}$ 
is about two times $\gamma$ for charm quark. Whereas, $\gamma_{HTL}$ for bottom is
about $40\%$ more than $\gamma$. We also observe
that this difference increases with the increase in temperature. 

\begin{figure}[h]      
\begin{center}                                                               
\includegraphics[scale=0.5]{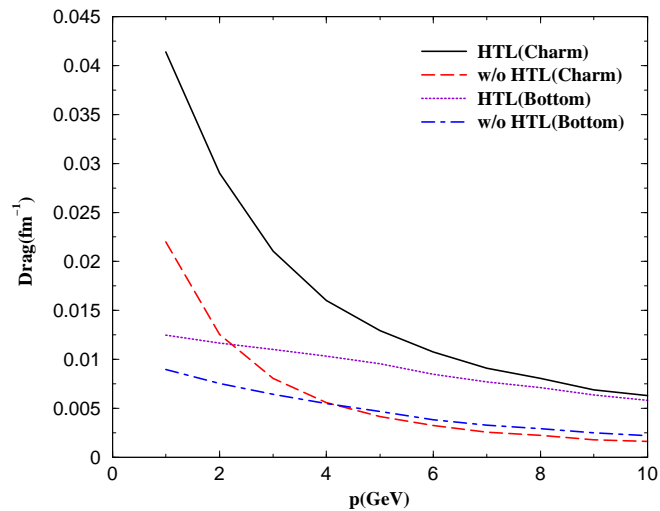}                                 
\end{center}                                                                 
\caption{ (Color online) Variation of drag of heavy quarks 
with momentum in a QGP bath of temperature 300 MeV.}
\label{fig3}                                                               
\end{figure}

The variation of $\gamma_{HTL}$ and  $\gamma$ 
with momentum is depicted in Fig.~\ref{fig3} for $T=300$ MeV.    
The $\gamma_{HTL}$ is greater than $\gamma$ for the entire momentum range considered here.
Again, drag being a measure of the HQs energy loss~\cite{braatenprd}, increase in drag results in 
more suppression of heavy flavours, which 
will have crucial consequences in understanding the heavy flavour suppression measured at RHIC and LHC energies. 
The momentum dependence of drag is distinctly affected if we consider the HTL resummation technique.
For a 5 GeV charm the  $\gamma_{HTL}$ is about three times $\gamma$ at $T=300$ MeV. 

\begin{figure}[h]      
\begin{center}                                                               
\includegraphics[scale=0.5]{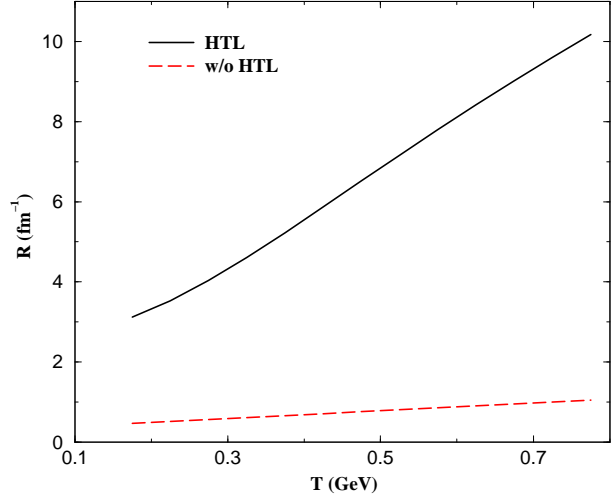}                                 
\end{center}                                                                 
\caption{(Color online) Temperature variation of the interaction rate of a charm quark  of 
momentum 3 GeV with QGP.}
\label{fig4}                                                               
\end{figure}

In Fig.~\ref{fig4} the interaction rate (see~\cite{rateq} for details) of the 
HQs with the QGP is shown as function of temperature.
The rate is increased when the HTL corrections in both the propagator and the
vertices are  taken into account.
The inclusion of the HTL approximated gluon propagator and vertices 
increases the likelihood of charm or bottom being equilibrated with the medium. 
The measured non-zero elliptic flow  of heavy flavours at RHIC (through the single
electron spectra originated from the semileptonic decays of the heavy flavoured mesons)  
and LHC energies indicate that the HQs in the QGP phase follow the collective motion 
of the background QGP, indicating thermalization of HQs.  
Therefore, the increase of the interaction rate (hence decrease in the interaction time scale) 
will have important implications in understanding the data on elliptic flow of heavy flavours.

\begin{figure}[h]      
\begin{center}                                                               
\includegraphics[scale=0.5]{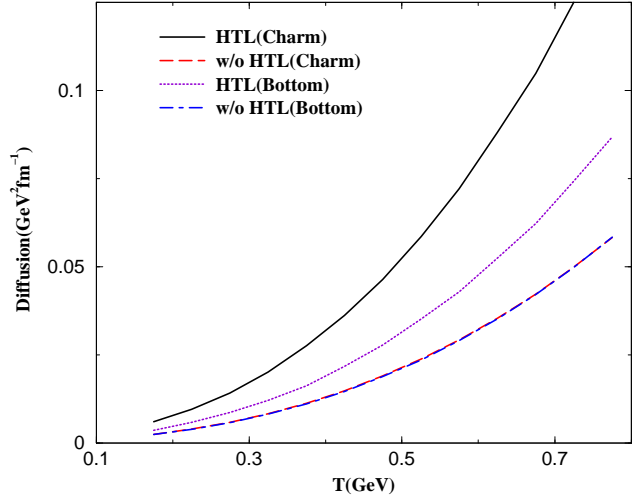}                                 
\end{center}                                                                 
\caption{ (Color online) Variation of diffusion of heavy quarks 
of momentum 1 GeV with temperature.}
\label{fig5}                                                               
\end{figure}

\begin{figure}[h]      
\begin{center}                                                               
\includegraphics[scale=0.5]{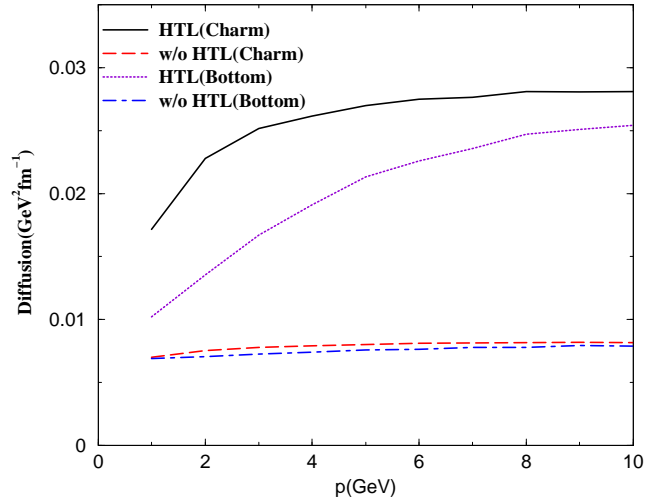}                                 
\end{center}                                                                 
\caption{ (Color online) Variation of diffusion of heavy quarks 
with momentum in a QGP bath of temperature 300 MeV.}
\label{fig6}                                                               
\end{figure}

In Figs.~\ref{fig5} and ~\ref{fig6} the diffusion coefficients $D_{HTL}$ and $D$
are plotted with $T$ (for $p=1$ GeV)  and $p$ (for $T=300$ MeV) respectively.   
At $T=300$ MeV, the magnitude of $D_{HTL}$ is almost 2.5 times $D$ in case of charm.
For beauty quark, the ratio of $D_{HTL}$ to $D$ is $\sim$ 1.4. 
The momentum dependence of diffusion is significantly modified too. A 5 GeV charm diffuses 3.3 times
more in momentum space when effective vertices and propagators are used. In case of 
bottom, the effect of effective propagators and vertices are less than that of charm.
These changes in drag and diffusion coefficients originate from the spectral modification of
the $t$-channel gluons due to its interaction with the thermal bath.
In the static limit $\pi_T\rightarrow 0$
and $\pi_L\rightarrow m_D^2$. The appearance of non-zero $\pi_T$ makes $\gamma_{HTL}$ larger than $\gamma$.
The inclusion of the vertex correction terms also introduces thermal fluctuations 
in the present formalism which, together with the resummed propagator, 
ultimately increases the magnitude of the drag and diffusion coefficients of HQ.

\begin{figure}[h]      
\includegraphics[scale=0.5]{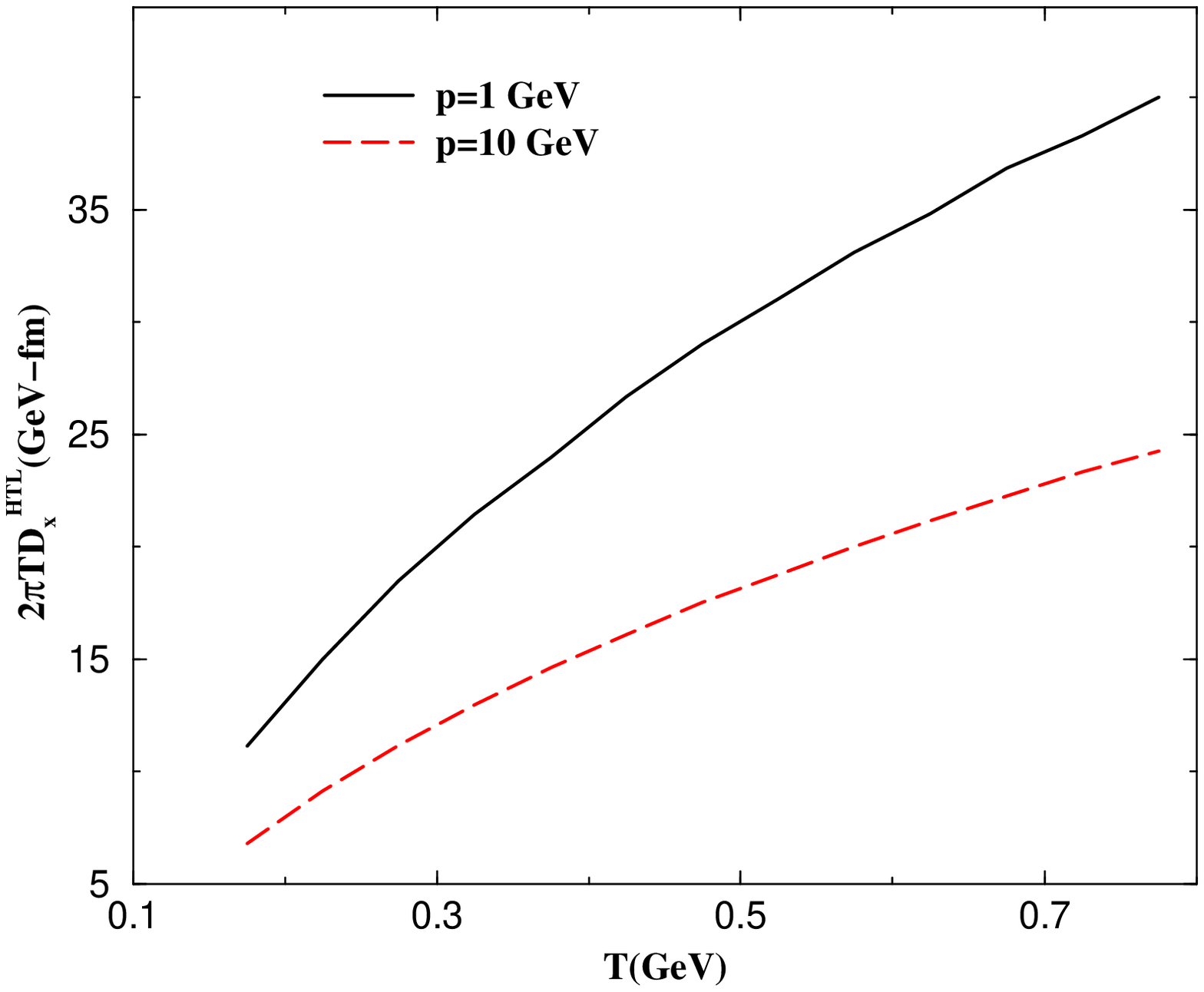}                                 
\caption{Comparison of co-ordinate space diffusion of charm quark with thermal de Broglie wavelength.}
\label{fig7}                                                               
\end{figure}

Now some comments on the magnitude of the values of $D_{HTL}$ and $D$ are in order 
here. The value of the diffusion coefficients in spatial co-ordinate, $D_x^{HTL}$ 
can be estimated from the value of drag by using the relation $D_x^{HTL}=T(m\gamma_{HTL})^{-1}$. 
In Fig.~\ref{fig7} we plot $D_x^{HTL}$ multiplied by the inverse of the thermal 
de  Broglie length, $\Lambda=1/(2\pi T)$. The results clearly indicate that 
the $D_x$ remains well above the quantum bound.  

\section{Summary and Conclusion:}
In summary, we have taken into account the HTL  modifications of the 
gluon spectral function and the $qqg$ and $ggg$ vertices  
in evaluating the drag and diffusion  coefficients of HQs propagating through the
QGP. The  deviations between $\gamma_{HTL}$ and $\gamma$ and $D_{HTL}$ and $D$ is found to be
substantial. The enhanced drag will result in higher suppression of the HQ momentum spectrum. 
The increase in drag will also enhance the chances of HQ getting equilibrated with
the bulk of the system. These results will have crucial consequences on the  
observables like nuclear suppressions, $R_{AA}(p_T)$ and elliptic flow, $v_2$ 
of heavy flavours measured at RHIC and LHC experiments. 
 

\section{Appendix: Calculating matrix elements from Hard Thermal Loop(HTL) Perturbation Theory}

\subsection{Symbols and Expressions we use:}
\begin{figure}[h]      
\includegraphics[scale=0.6]{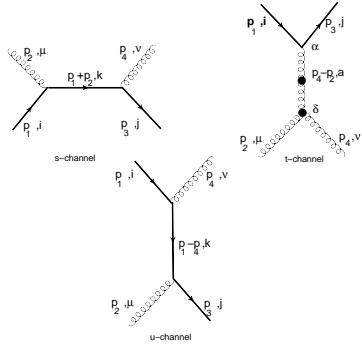}                                 
\caption{$Qg\rightarrow Qg$ Feynman diagrams}
\label{fig8}                                                               
\end{figure}

In this appendix we evaluate the following matrix elements for the processes 
$Qg\rightarrow Qg$  (Fig.~\ref{fig8}) and $Qq\rightarrow Qq$  
(Fig.~\ref{fig11}) in a thermal medium applying HTL approximations. 
First, we define the following useful quantities~\cite{bellac}
required to write down the gluon propagator 
in thermal medium. Let $u_{\mu}$ be
the fluid four-velocity, with normalization condition $u^\mu_\mu=1$. Then any four-vector 
$P^\mu$ can be decomposed into components parallel and perpendicular to the fluid velocity:
\bea
\omega=P.u\nn\\
\tilde{P}_{\mu}=P_{\mu}-u_{\mu}(P.u)\nn\\
\label{eq1}
\eea
where
\bea
P^2=\omega^2-p^2 \nn\\
\tilde{P}^2=-p^2
\label{eq2}
\eea
Eqs. \ref{eq1} and \ref{eq2} are valid in the local rest frame of fluid, i.e. in a frame where
$u=(1,\vec{0})$. Similarly a tensor orthogonal to $u_{\mu}$ can be defined as,
\be
\tilde{g}_{\mu \nu}=g_{\mu \nu}-u_{\mu}u_{\nu}
\label{eq3}
\ee
The longitudinal and transverse projection tensors $\mathcal{P}_L^{\mu\nu}$ 
and  $\mathcal{P}_T^{\mu\nu}$ respectively are defined as~\cite{heisel}
\bea
\mathcal{P}^{\mu\nu}_L=-\frac{1}{P^2 p^2} (\omega P^{\mu}-P^2 u^{\mu})(\omega P^{\nu}-P^2 u^{\nu})
\label{eq5}
\eea
\bea
\mathcal{P}^{\mu\nu}_T=\tilde{g}_{\mu \nu}+\frac{\tilde{P}_{\mu}\tilde{P}_{\nu}}{p^2}
\label{eq6}
\eea
which are orthogonal to $P^{\mu}$ as well as to each other, {\it  i.e.}
\be
P_{\mu}\mathcal{P}^{\mu\nu}_L=P_{\mu}\mathcal{P}^{\mu\nu}_T=\mathcal{P}^{\mu}_{L\nu}\mathcal{P}^{\nu\rho}_T=0
\label{eq7}
\ee
But,
\be
\mathcal{P}^{\mu\rho}_i\mathcal{P}_{i\nu\rho}=\mathcal{P}_{i\nu}^{\mu}~~~,\mathrm{i}=\mathrm{L, T}
\label{eq8}
\ee
The transverse and longitudinal self-energies of gluon at non-zero temperature are given by:
\bea
\Pi_T(P)=(1-x^2)\pi_L(x),~~~~\Pi_L(P)=\pi_T(x)
\label{eq10}
\eea
respectively, where $x=\omega/p$ and scaled self-energies $\pi_T$ and $\pi_L$ are given by~\cite{bellac}, 

\bea
\pi_T(x)=
m_D^2\left[\frac{x^2}{2}+\frac{x}{4}(1-x^2)ln \left(\frac{1+x}{1-x}\right) 
-i\frac{\pi}{4}x(1-x^2)\right]\nn\\
\eea
and
\bea
\pi_L(x)=m_D^2\left[1-\frac{x}{2}ln (\frac{1+x}{1-x})+i \frac{\pi}{2}x\right]\nn\\
\label{eq11}
\eea
respectively. Non-zero real and imaginary parts of the self-energies corresponds 
to the shift of the pole of the propagator and to  the different physical processes 
take place in the  medium.
With the help of the quantities defined above we can now write down  
the gluon propagator with momentum $P$ using 
Dyson-Schwinger equation:
\be
\Delta^{\mu\nu}=\frac{\mathcal{P}_T^{\mu\nu}}{-P^2+\Pi_T}+\frac{\mathcal{P}_L^{\mu\nu}}{-P^2+\Pi_L}
+(\alpha-1)\frac{P^{\mu}P^{\nu}}{P^2}
\label{eq4}
\ee
where $\alpha$ is a gauge-fixing parameter taken to be unity in this literature.


\subsection{Calculating $Qg\rightarrow Qg$ Matrix Element:}

This process contains three Feynman diagrams corresponding to the channels s, t and u. 
Since heavy quarks are not thermalized, we use bare  HQ
propagators as well as bare HQ-gluon vertex for s channel and u channel diagrams. 
Consequently, we use naive perturbation theory results~\cite{sve,combridge}
for $|M_s|^2,~|M_u|^2$ and cross-term $M_sM_u^*$. On the other hand, we have to use effective propagator 
as well as effective three-gluon vertex for t channel diagram. Hence, $|M_t|^2$ and cross-terms $|M_sM_t^*|$ as well
as $|M_uM_t^*|$ are drastically different from their $T=0$ counterparts. We write down 
$M_s,~M_t$ ~and~$M_u$ for the process under discussion.

\be
-iM_s=\overline{u}(P_3)(-ig\gamma^{\nu}t^b_{jk})i\frac{P_1\sls+P_2\sls }{s-m^2}(-ig\gamma^{\mu}t^c_{ki})u(P_1)
\epsilon_{\mu}\epsilon_{\nu}^*
\label{eq12}
\ee
\be
-iM_u=\overline{u}(P_3)(-ig\gamma^{\mu}t^c_{jk})i\frac{P_1\sls-P_4\sls }{u-m^2}(-ig\gamma^{\nu}t^b_{ki})u(P_1)
\epsilon_{\mu}\epsilon_{\nu}^*
\label{eq13}
\ee
\be
-iM_t=\overline{u}(P_3)(-ig\gamma^{\alpha}t^a_{ji})u(P_1)(-i\Delta_{\alpha\delta})
gf_{abc} \Gamma^{\mu\delta\nu}
\epsilon_{\mu}\epsilon_{\nu}^*
\label{eq14}
\ee
Here, $s=(P_1+P_2)^2,~t=(P_4-P_2)^2~\mathrm{and}~u=(P_1-P_4)^2$ are Mandelstam variables.
Now, according to the requirement of the gauge invariance 
we have used both three-gluon ($ggg$) effective HTL 
vertex ($\Gamma^{\mu\delta\nu}$) and HTL resummed gluon propagator ($\Delta_{\alpha\delta}$). 
We note that $ggg$ effective vertex $\Gamma_{\mu \delta \nu}$ is given by two parts:
(a) the $T=0$ $ggg$ vertex ($\mathcal{C}_{ \mu \delta \nu}$)
and (b) the one-loop HTL correction to (a), $\delta \Gamma_{\mu\delta\nu}$.
So we can write:
\be
\Gamma_{\mu \delta \nu}=\mathcal{C}_{ \mu \delta \nu}
+\delta \Gamma_{\mu\delta\nu}
\label{eq15}
\ee

where $\mathcal{C}^{\mu\delta\nu}=[(2P_4-P_2)^\mu g^{\delta\nu}+(-P_4-P_2)^\delta g^{\mu\nu}
+(2P_2-P_4)^\nu g^{\mu\delta}]$ is three-gluon vertex at zero temperature . 

\begin{figure}[h]      
\begin{center}                                                               
\includegraphics[scale=0.9]{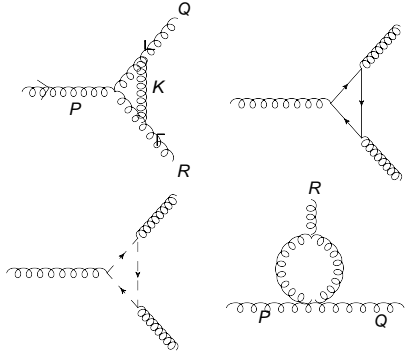}                                 
\end{center}                                                                 
\caption{One-loop Feynman graphs which contribute to HTL in $ggg$ vertex. The solid line is for fermions. 
The dotted line is for ghost.}
\label{fig9}                                                               
\end{figure}

 
The diagrams which contribute to HTL correction to $ggg$ vertex are given in Fig.\ref{fig9}.
The expression for HTL $ggg$ vertex correction is :
\bea
\delta \Gamma_{\mu\delta\nu}=2 m_D^2 \int \frac{d\Omega}{4\pi} \hat{K}_{\mu}\hat{K}_{\delta}\hat{K}_{\nu} 
\times \left[\frac{i\omega_r}{(P.\hat{K})(R.\hat{K})}-\frac{i\omega_q}{(P.\hat{K})(Q.\hat{K})}\right]
\label{eq19}
\eea 
The momenta $P,~Q,~R$ are defined in Fig. \ref{fig10} with $P+Q+R=0$. 
For Simplicity we assume three momentum 
transfer, $\vec{p}$, to be zero~\cite{bellac}. In this approximation we can simplify Eq. \ref{eq19} into:
\bea
\delta \Gamma_{\mu\delta\nu}= -2 m_D^2 \int \frac{d\Omega}{4\pi} \hat{K}_{\mu}\hat{K}_{\delta}\hat{K}_{\nu}
\frac{|\vec{q}|cos{\theta}}{R.\hat{K}Q.\hat{K}},
\label{eq20}
\eea
where $cos{\theta}$ is the angle $\vec{q}$ (considered to be along z-axis) makes with $\hat{K}$.

The complex-conjugate of Eq.~\ref{eq20} can be written as:
\bea
\delta \Gamma_{\mu\delta\nu}^*&=& -2 m_D^2 \int \frac{d\Omega}{4\pi} \hat{K}_{\mu}\hat{K}_{\delta}\hat{K}_{\nu}
\frac{|\vec{q}|cos{\theta}}{R.\hat{K}Q.\hat{K}}\nn\\
&=&\delta \Gamma_{\mu\delta\nu}
\label{eq21}
\eea

\section*{Calculating $t$-channel diagram}
The t-channel diagram for the process $Qg\rightarrow Qg$ has the following
amplitude square:

\bea
\frac{8}{g^4}\overline{M_tM_t^*}&=&[4(m^2-P_1.P_3)g^{\alpha\alpha^{\prime}}
+4P_1^{\alpha}P_3^{\alpha'}
+4P_3^{\alpha}P_1^{\alpha'}]\nn\\
&&\times \Delta_{\alpha\delta}\Delta^*_{\alpha^{\prime}\delta^{\prime}}
\underbrace{\Gamma^{\mu \delta \nu}\Gamma^{*\mu' \delta' \nu'} g_{\mu\mu^{\prime}}
g_{\nu\nu^{\prime}}}_{\Gamma^{\delta}\cdot \Gamma^{*\delta'}}\nn\\
&=&[\underbrace{4(m^2-P_1.P_3)g^{\alpha\alpha^{\prime}}
\Delta_{\alpha\delta}\Delta^*_{\alpha^{\prime}\delta^{\prime}}}_A
+\underbrace{4P_1^{\alpha}P_3^{\alpha^{\prime}}
\Delta_{\alpha\delta}\Delta^*_{\alpha^{\prime}\delta^{\prime}}}_B
+\underbrace{4P_3^{\alpha}P_1^{\alpha^{\prime}}
\Delta_{\alpha\delta}\Delta^*_{\alpha^{\prime}\delta^{\prime}}}_C]\nn\\
&&\times
\left(\Gamma^{\delta}\cdot \Gamma^{*\delta'}\right),
\label{eq22}
\eea
where $m$ is the mass of Heavy Quark (HQ).
The term $\Gamma^{\delta}\cdot \Gamma^{*\delta'}$ is actually given by the following terms:
\bea
\Gamma^{\delta}\cdot \Gamma^{*\delta'}=
\Gamma^{\mu \delta \nu} \Gamma_{\mu\nu}^ {\delta'}\sim
\mathcal{O}(m_D^0)+\mathcal{O}(m_D^2)+\mathcal{O}(m_D^4),
\label{eq23}
\eea
We are left with the contributions from (a) product of uncorrected vertices $(\mathcal{O}(m_D^0))$, 
(b) product of corrected and uncorrected vertices ($\mathcal{O}({m_D}^2)$) and (c) that of HTL corrections 
$(\mathcal{O}(m_D^4))$. Calculation of part (a) can be performed
by taking explicitly the form of the HTL resummed gluon propagator. 
We are not writing down the full expression for part (a) simply because it is too long. However, 
one can evaluate part (b) as well as (c) with the assumption that the three momentum transfer, $\vec{p}$ is negligibly small. 
Here, we will illustrate the calculation of the terms of Eq. \ref{eq22} having contributions of 
($\mathcal{O}({m_D}^2)$) and ($\mathcal{O}({m_D}^4)$).

It is evident from Eq. \ref{eq22} that we need to evaluate the quantity $\Gamma^{\delta}\cdot \Gamma^{*\delta'}$
as discussed below:

\begin{figure}[h]      
\begin{center}                                                               
\includegraphics[scale=0.7]{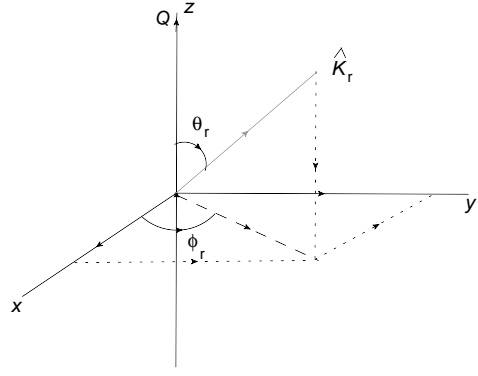}                                 
\end{center}                                                                 
\caption{Choice of angles in spherical polar co-ordinate}
\label{fig10}                                                               
\end{figure}                                                                 

\bea
\Gamma^{\delta}\cdot \Gamma^{*\delta'}&=&\underbrace{
\mathcal{C}'^{\delta'}_{\mu\nu}\mathcal{C}'^{\delta\mu\nu}}_{\sharp 1}
+\underbrace{2 \mathcal{C}'^{\delta'}_{\mu\nu}\delta\Gamma^{\mu \delta \nu}}_{\sharp 2}
+\underbrace{\delta\Gamma^{\mu \delta \nu} \delta \Gamma_{\mu\nu}^ {\delta'}}_{\sharp 3}\nn\\
&=&\mathcal{C}'^{\delta'}_{\mu\nu}\mathcal{C}'^{\delta\mu\nu}+
2 \left[-2m_D^2 \int \frac{d\Omega}{4\pi} \hat{K}^{\mu}\hat{K}^{\delta}\hat{K}^{\nu}
\frac{|\vec{q}|cos{\theta}}{R.\hat{K}Q.\hat{K}}\mathcal{C}'^{\delta'}_{\mu\nu}\right]\nn\\
&&+4m_D^4 \int\frac{d\Omega_1}{4\pi}\frac{d\Omega_2}{4\pi} (\hat{K}_1\cdot\hat{K}_2)^2
\hat{K}_1^{\delta}\hat{K}_2^{\delta'}\nn\\
&&\times \frac{|\vec{q}|^2 cos\theta_1cos\theta_2}{Q.\hat{K}_1R.\hat{K}_1Q.\hat{K}_2R.\hat{K}_2} ~,
\label{eq24}
\eea
where $\Omega_1(\Omega_2)$ is the solid angle $\hat{K}_1(\hat{K}_2)$ makes with Q(chosen to be along z-axis)
(Fig.\ref{fig10}). From three-momentum conservation at the three-gluon vertex we get,
$
\vec{p}+\vec{q}+\vec{r}=0\,\,\,\,
\vec{p}=0~~\Rightarrow\vec{q}=-\vec{r}~~;~~|\vec{q}|=|\vec{r}|$
Hence $\vec{r}$ is aligned along the negative z axis. Now, choosing $\hat{K}_r=(-i,\hat{k}_r),~[r=1,2]$, 
a light-like unit vector, we can write three-unit vectors $\hat{k}_r$ as below:
\bea
\hat{k}_r=sin\theta_r cos\phi_r \hat{i}+sin\theta_r sin\phi_r \hat{j}+cos\theta_r\hat{k}
\label{eq26}
\eea
in the spherical polar co-ordinate.
\bea
(\hat{K}_1\cdot\hat{K}_2)^2&=&(-1+\hat{k}_1\cdot\hat{k}_2)^2\nn\\
&=&sin^2\theta_1sin^2\theta_2 cos^2(\phi_1-\phi_2)+cos^2\theta_1cos^2\theta_2 \nn\\
&&+sin\theta_1sin\theta_2 cos(\phi_1-\phi_2)cos\theta_1cos\theta_2\nn\\
&&-2(sin\theta_1sin\theta_2 cos(\phi_1-\phi_2)+cos\theta_1cos\theta_2)+1\nn\\
Q.\hat{K_r}&=&-iQ_4+|\vec{q}|cos\theta_r ~~~~~(Euclidean)\nn\\ 
R.\hat{K_r}&=&-iR_4-|\vec{r}|cos\theta_r
\label{eq27}
\eea
So the integrand in Eq. \ref{eq24} is, now, entirely in terms of $\theta_r$ and $\phi_r$.
Having done so, we will find out an analytic expression for right hand side of Eq. \ref{eq22}.

\vskip 0.2in

\section*{\large Calculation of $|\overline{M_t}|^2$:}

\vskip 0.2in
\noi Equation \ref{eq22} can be written in the following way,
\bea
\frac{8}{g^4}\overline{M_tM_t^*}&=&[A+B+C]\times [\sharp 1+\sharp 2+\sharp 3],
\label{eq28}
\eea
where A, B, C and $\sharp 1$, $\sharp 2$ and $\sharp 3$ are already defined. Now let us 
evaluate the products one by one. 

\vskip 0.2in

\noi {\bf \large Calculation of $(A+B+C)\times \sharp 1$:}

\vskip 0.2in

This is the part of $|M_t|^2$ which involves propagator correction only and no vertex correction.
Since heavy quarks may not thermalize , we use $T=0$ results for $\overline{|M|}_s^2$
, $\overline{|M|}_u^2$, $\overline{M_sM_u^*}$ \cite{combridge} in our formalism. Results for $\overline{|M|}_t^2$,
$\overline{M_sM_t^*}$, $\overline{M_uM_t^*}$ has a quite long expression which we have not written here.

\vskip 0.2in

\noi {\bf \large Calculation of $A\times \sharp 2$:}

\vskip 0.2in

This calculation can be done following the procedure as delineated below:

\bea
A\times \sharp 2 &=&4(m^2-P_1.P_3)g^{\alpha\alpha'}
\Delta_{\alpha\delta}\Delta^*_{\alpha^{\prime}\delta^{\prime}}
\times 2 \left[-2m_D^2 \int \frac{d\Omega}{4\pi} \hat{K}^{\mu}\hat{K}^{\delta}\hat{K}^{\nu}
\frac{|\vec{q}|cos{\theta}}{R.\hat{K}Q.\hat{K}}\mathcal{C}'^{\delta'}_{\mu\nu}\right]\nn\\
&=&-8tm_D^2\left[\frac{\mathcal{P}_{T\delta \delta '}}{|t-\Pi_T|^2}+\frac{\mathcal{P}_{L\delta \delta '}}
{|t-\Pi_L|^2}\right]
\times \left[ \int \frac{d\Omega}{4\pi} \hat{K}^{\mu}\hat{K}^{\delta}\hat{K}^{\nu}
\frac{|\vec{q}|cos{\theta}}{R.\hat{K}Q.\hat{K}}\mathcal{C}'^{\delta'}_{\mu\nu}\right]
\label{eq29}
\eea

\noi Writing the entire expression for transverse and longitudinal projection operators and 
$\mathcal{C'}^{\delta'}_{\mu\nu}$we find that the terms we need to evaluate in Euclidean space contains the following: 
(a) $g_{\delta\delta'}\hat{K}^{\delta}\hat{K}^{\delta'}$
(b) $u_{\delta}u_{\delta'}\hat{K}^{\delta}\hat{K}^{\delta'}$,
(c)$(P.\hat{K}-\omega u.\hat{K})$ and
(d)$(\omega P.\hat{K}-P^2 u.\hat{K})$, save the coefficients. 
(a) is zero because $\hat{K}$ is light-like. (b) is -1. (c) and (d) both give zero by virtue of
the assumption of vanishing three-momentum transfer. Finally, we get

\bea
A\times \sharp 2 &=&-\frac{16tm_D^2}{|t-\Pi_T|^2}\left[2-a_1log\left(\frac{a_1+1}{a_1-1}\right)\right]~,
\label{eq30}
\eea 
where $a_1=\frac{Q_0}{|\vec{q}|}$, after analytic continuation to Minkowski space.

\vskip 0.2in

\noi {\bf \large Calculation of $(B+C)\times \sharp 2$:}

\vskip 0.2in

Can be shown to be zero if we average over the directions of heavy quark momenta.

\vskip 0.2in

\noi {\bf \large Calculation of $A\times \sharp 3$:}

\vskip 0.2in

\noi The calculation is depicted below:

\bea
A\times \sharp 3&=& 4(m^2-P_1.P_3)g^{\alpha\alpha'}
\Delta_{\alpha\delta}\Delta^*_{\alpha^{\prime}\delta^{\prime}}
\times 4m_D^4 \int\frac{d\Omega_1}{4\pi}\frac{d\Omega_2}{4\pi} (\hat{K}_1\cdot\hat{K}_2)^2
\hat{K}_1^{\delta}\hat{K}_2^{\delta'}
\times \frac{|\vec{q}|^2 cos\theta_1cos\theta_2}{Q.\hat{K}_1R.\hat{K}_1Q.\hat{K}_2R.\hat{K}_2}
\label{eq31}
\eea
Eq. \ref{eq31} involves the following,

\vskip 0.2in

\noi \textbullet   
\bea
g_{\delta\delta'}\left(\delta\Gamma^{\mu\delta\nu}\cdot \Gamma^{*\delta'}_{\mu\nu}\right)=
4m_D^4 \int\frac{d\Omega_1}{4\pi}\frac{d\Omega_2}{4\pi} (\hat{K}_1\cdot\hat{K}_2)^3
\frac{|\vec{q}|^2 cos\theta_1cos\theta_2}{Q.\hat{K}_1R.\hat{K}_1Q.\hat{K}_2R.\hat{K}_2}
\label{eq32}
\eea
After expanding $(\hat{K}_1\cdot\hat{K}_2)^3$, we retain only those terms which will yield non-zero
contribution in Eq. \ref{eq32}. The final result for part (a) gives:


\bea
g_{\delta\delta'}\left(\delta\Gamma^{\mu\delta\nu}\cdot \Gamma^{*\delta'}_{\mu\nu}\right)&=&
\frac{3m_D^4}{|\vec{q}|^2}\left[2-a_1 log\left(\frac{a_1+1}{a_1-1}\right)\right]^2\nn\\
&&+\frac{m_D^4}{|\vec{q}|^2}\left[\frac{2}{3}+2a_1^2-a_1^3 log\left(\frac{a_1+1}{a_1-1}\right)\right]^2\nn\\
&&+\frac{3m_D^4}{2|\vec{q}|^2}\left[\frac{4}{3}-2a_1^2-(a_1-a_1^3) log\left(\frac{a_1+1}{a_1-1}\right)\right]^2\nn\\
\label{eq33}
\eea
where we have used 

\bea
\int\frac{d\Omega}{4\pi} \frac{cos^{(2n+1)}\theta}{(-ia+cos\theta)(-ia-cos\theta)}&=&0  ~~[n\in \mathcal{I}^+]\nn\\
\int_0^{2\pi} cos(\phi_1-\phi_2)d\phi_1d\phi_2&=&0 
\label{eq34}
\eea
and have utilized known results for 

\bea
\int\frac{d\Omega}{4\pi} \frac{cos^{2n}\theta}{(-ia+cos\theta)(-ia-cos\theta)}  ~~[n\in \mathcal{I}^+]
\label{eq35}
\eea
and 

\vskip 0.2in

\noi \textbullet 
\bea
u_{\delta}u_{\delta'}\left(\delta\Gamma^{\mu\delta\nu}\cdot \Gamma^{*\delta'}_{\mu\nu}\right)
=
-\frac{2m_D^4}{|\vec{q}|^2}\left[2-a_1 log\left(\frac{a_1+1}{a_1-1}\right)\right]^2
\label{eq36}
\eea

The rest are zero as:

\vskip 0.2in

\noi \textbullet 
\bea
P.\hat{K}-\omega u.\hat{K}=-i\omega +i\omega=0
\label{eq37}
\eea
and

\vskip 0.2in

\noi \textbullet 

\bea
\omega P.\hat{K}-P^2 u.\hat{K}&=&\omega P.\hat{K}-\omega^2 u.\hat{K} \nn\\
&=&\omega(-i\omega +i\omega)\nn\\
&=&0 \nn\\
&&[because P^2=\omega^2,~in~the~approximation~\vec{p}=0]
\label{eq38}
\eea

\vskip 0.2in

\noi {\bf \large Calculation of B $\times \sharp 3$:}

\vskip 0.2in
This term has vanishing contribution  if average over the directions of $P_1$ or
$P_3$ are taken. Same is true for the term 
$C\times \sharp 3$.

\vskip 0.5 in
\noi {\large \textbullet~~~} Hence the expression for $|M_t|^2$  due to both the vertex and propagator corrections becomes:

\bea
\frac{|t-\Pi_T|^2}{2t}\frac{8}{g^4}\overline{M_tM_t^*}
&=&\frac{m_D^4}{|\vec{q}|^2}\left[2-a_1 log\left(\frac{a_1+1}{a_1-1}\right)\right]^2
+\frac{m_D^4}{|\vec{q}|^2}\left[\frac{2}{3}+2a_1^2-a_1^3 log\left(\frac{a_1+1}{a_1-1}\right)\right]^2\nn\\
&&+\frac{3m_D^4}{2|\vec{q}|^2}\left[\frac{4}{3}-2a_1^2-(a_1-a_1^3) log\left(\frac{a_1+1}{a_1-1}\right)\right]^2
-8 m_D^2 \left(2-2\frac{a_1}{2}log\left(\frac{a_1+1}{a_1-1}\right)\right)
\label{eq39}
\eea

\vskip 0.2in

\section*{\large Contribution of $M_sM_t^*$:}

\vskip 0.2in

\noi $M_sM_t^*$ can be calculated applying all the assumptions and techniques already discussed. So we can 
directly write down the results (with vertex and propagator corrections).
\bea
&|t-\Pi_T|^2&\cdot\frac{16 }{g^4}(s-m^2)Re\overline{M_s M_t^*}=
-8tm_D^2\left[\left(1-\frac{a_1}{2}log\left|\frac{a_1+1}{a_1-1}\right|\right)(t-Re\Pi_T)\right]
\label{eq40}
\eea
Similar procedure may be followed to obtain the corresponding expressions for $M_uM_t^*$.

\bea
&|t-\Pi_T|^2&\cdot\frac{16 }{g^4}(u-m^2)Re\overline{M_u M_t^*}=
8tm_D^2\left[\left(1-\frac{a_1}{2}log\left|\frac{a_1+1}{a_1-1}\right|\right)(t-Re\Pi_T)\right]
\label{eq41}
\eea

\subsection{$Qq\rightarrow Qq$ Matrix Element from HTL approximation:} 

\begin{figure}[h]      
\includegraphics[scale=0.6]{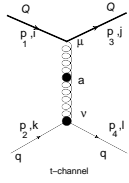}                                 
\caption{$Qq\rightarrow Qq$ Feynman diagram. Bold lines are for heavy quarks(HQ).}
\label{fig11}                                                               
\end{figure}
\begin{figure}[h]      
\begin{center}                                                               
\includegraphics[scale=0.9]{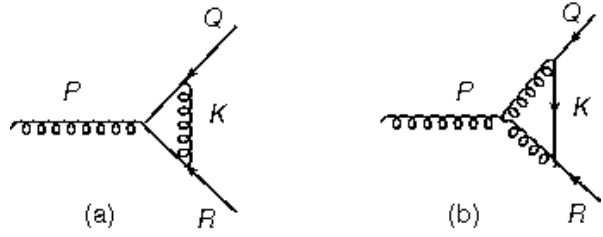}                                 
\end{center}                                                                 
\caption{Feynman diagrams contributing to HTL correction to $qqg$ vertex upto one loop}
\label{fig12}                                                               
\end{figure}

       
\noi The effective propagator and the effective vertex will be denoted by solid circles. We can write the amplitude 
for $Qq\rightarrow Qq$ in Feynman Gauge($\alpha=1$) from Fig. \ref{fig11} as,
\bea
-iM_t= \overline{u}(P_3)(-ig\gamma^{\mu}t^a_{ji})u(P_1)\left[-i\Delta_{\mu\nu}\right]\nn\\
\overline{u}(P_4)(-ig\Gamma^{\nu}t^a_{lk})u(P_2)
\label{eq42}
\eea
$i,j,k,l~~(i\neq j,~k\neq l)$ are quark colours and 
`a' is the colour of intermediary  gluon with polarizations $\mu,\nu$.
The term $\Gamma^{\nu}$ denotes the HTL vertex (Fig.\ref{fig12}) 
correction term upto one loop and is given by the following expression.

\bea
\Gamma^{\nu}=\gamma^{\nu}+m_{f}^{2}\int\frac{d\Omega}{4\pi}\frac{\hat{k}^{\nu}
\hat{k\sls}}{(Q.\hat{k})(R.\hat{k})}~,
\label{eq43}
\eea 
where $m_f=gT/\sqrt{6}$ is the thermal mass of fermions.
After squaring and averaging over spin and colour and using Eq. \ref{eq4} we get
\bea
\sum|M_{Qq}|^{2}=g^4 C_{Qq}\left[4(m^{2}-P_{1}.P_{3})g^{\mu \mu'}+4P_{1}^{\mu}P_{3}^{\mu'}
+4P_{3}^{\mu}P_{1}^{\mu'}\right]\Delta_{\mu\nu}\Delta^*_{\mu'\nu'}Tr\left[P_{4}\sls 
\Gamma^\nu P_{2}\sls \Gamma^{\nu'}\right]
\label{eq44}
\eea 
First we calculate the trace involving the vertex correction term.
\bea
Tr\left[P_{4}\sls \Gamma^\nu P_{2}\sls \Gamma^{\nu'}\right]=Tr\left[P_{4}\sls \gamma^{\nu}P_{2}\sls \gamma^{\nu'}\right]
-m_{f}^2\int Tr\left[P_{4}\sls \gamma^\nu P_{2}\sls \hat{k\sls}\right]\frac{\hat{k}^{\nu'}}{(P_{4}.\hat{k})(P_{2}.\hat{k})}
\frac{d\Omega}{4\pi}\nn\\ -m_{f}^2\int Tr\left[P_{4}\sls \hat{k\sls}P_{2}\sls \gamma^{\nu'}\right]
\frac{\hat{k}^{\nu}}{(P_{4}.\hat{k})(P_{2}.\hat{k})}\frac{d\Omega}{4\pi}\nn\\ +m_{f}^4\int Tr\left[P_{4}\sls \hat{k_1\sls}
P_{2}\sls \hat{k_2\sls}\right]\frac{\hat{k_1}^{\nu}}{(P_{4}.\hat{k_1})(P_{2}.\hat{k_1})}
\frac{\hat{k_2}^{\nu'}}{(P_{4}.\hat{k_2})(P_{2}.\hat{k_2})}\frac{d\Omega_1}{4\pi}\frac{d\Omega_2}{4\pi}
\label{eq45}
\eea
After having calculated the trace and performed the relevant integration we arrive at the final result which 
constitutes of two parts: the first part comes solely from the correction due to HTL propagator and the rest
is attributed to corrections due to HTL propagator as well as HTL vertex upto one loop approximation.
The first part due to propagator correction only is given by:
\bea
\frac{\sum\left|M_{Qq}\right|^2}{4 C_{Qq} g^4}&=&2\frac{P_4.\mathcal{P}_T.P_3 P_2.\mathcal{P}_T.P_1}{\left(t-\Pi_T\right)^2}
+2\frac{P_4.\mathcal{P}_L.P_3 P_2.\mathcal{P}_L.P_1}{\left(t-\Pi_L\right)^2}
+2\frac{P_4.\mathcal{P}_T.P_1 P_2.\mathcal{P}_T.P_3}{\left(t-\Pi_T\right)^2}
+2\frac{P_4.\mathcal{P}_L.P_1 P_2.\mathcal{P}_L.P_3}{\left(t-\Pi_L\right)^2}\nn\\
&+&2A \frac{P_4.\mathcal{P}_L.P_3 P_2.\mathcal{P}_T.P_1+P_4.\mathcal{P}_T.P_3 P_2.\mathcal{P}_L.P_1}
{\left(t-\Pi_T\right)^2 \left(t-\Pi_L\right)^2}
+2A\frac{P_4.\mathcal{P}_L.P_1 P_2.\mathcal{P}_T.P_3+P_4.\mathcal{P}_T.P_1 P_2.\mathcal{P}_L.P_3}
{\left(t-\Pi_T\right)^2 \left(t-\Pi_L\right)^2}\nn\\
&-&2P_4.P_2\left[\frac{P_3.\mathcal{P}_T.P_1}{\left(t-\Pi_T\right)^2}
+\frac{P_3.\mathcal{P}_L.P_1}{\left(t-\Pi_L\right)^2}\right]
-2P_3.P_1\left[\frac{P_4.\mathcal{P}_T.P_2}{\left(t-\Pi_T\right)^2}
+\frac{P_4.\mathcal{P}_L.P_2}{\left(t-\Pi_L\right)^2}\right]\nn\\
&+&P_3.P_1P_4.P_2\left[\frac{2}{\left(t-\Pi_T\right)^2}+\frac{1}{\left(t-\Pi_L\right)^2}\right]
+m^2\left[2\frac{P_4.\mathcal{P}_T.P_2}{\left(t-\Pi_T\right)^2}
+2\frac{P_4.\mathcal{P}_L.P_2}{\left(t-\Pi_L\right)^2}\right]\nn\\
&-&m^2\left[2\frac{P_4.P_2}{\left(t-\Pi_T\right)^2}+\frac{P_4.P_2}{\left(t-\Pi_L\right)^2}\right]
\label{eq46}
\eea
where $C_{Qq}=\frac{2}{9}$ is the color factor, $ t=(P_1-P_3)^2$, 
$A=t^2-t(Re\Pi_T+Re\Pi_L)+Re\Pi_T\Pi_L^*$ and we have used the following relations.

\bea
\Delta^{\mu\rho}\Delta^{*\nu}_\rho&=&\frac{\mathcal{P}_T^{\mu\nu}}{\left(t-\Pi_T\right)^2}
+\frac{\mathcal{P}_L^{\mu\nu}}{\left(t-\Pi_L\right)^2}\nn\\
\left|\Delta\right|^2&=&\Delta^{\mu\nu}\Delta^*_{\nu_\mu}
=\frac{2}{\left(t-\Pi_T\right)^2}+\frac{1}{\left(t-\Pi_L\right)^2}\nn\\
\label{eq47}
\eea

Using eqs. \ref{eq5}, \ref{eq6} we can show that

\bea
P_1.\mathcal{P}_L.P_2=P_3.\mathcal{P}_L.P_4=
P_4.\mathcal{P}_L.P_1=P_2.\mathcal{P}_L.P_3
\label{eq48}
\eea
where all the calculations are done in the rest frame of fluid element. 
Therefore, the terms due to both HTL approximated vertices and propagator are:
\bea
\frac{\sum |M_{Qq}|_{v.c}^2}{g^4C_{Qq}}&=&-32t\frac{m_{f}^2}{\left|t-\Pi_T\right|^2}+\frac{4t^2m_{f}^2}{\left|t-\Pi_T\right|^2
|\vec{q}|^2a_1}ln\left(\frac{a_1+1}{a_1-1}\right)+16ta_1\frac{m_{f}^2}{\left|t-\Pi_T\right|^2}ln\left(\frac{a_1+1}{a_1-1}\right)\nn\\
&+& \frac{4tm_{f}^4}{\left|t-\Pi_T\right|^2|\vec{q}|^2}\left[{2-a_{1}ln\left(\frac{a_1+1}{a_1-1}\right)}^2\right]+
\frac{2t^2m_{f}^4}{\left|t-\Pi_T\right|^2|\vec{q}|^4}\left[\frac{a_1}{2}ln\left(\frac{a_1+1}{a_1-1}\right)
-\frac{1}{2a_1}ln\left(\frac{a_1+1}{a_1-1}\right)-1\right]^2\nn\\
&+&\frac{4t^2m_{f}^4}{\left|t-\Pi_T\right|^2|\vec{q}|^4}\left[1-\frac{a_1}{2}ln\left(\frac{a_1+1}{a_1-1}\right)\right]^2
\label{eq49}
\eea 
where, $a_1=Q_0/|\vec{q}|$ in Minkowski space.

\noindent{\bf Acknowledgement:} SM and TB are supported by DAE, Govt. of India. Fruitful discussions with
Sourav Sarkar, Najmul Haque and M. G. Mustafa are acknowledged. 


\end{document}